\documentclass[fleqn,usenatbib]{mnras}

\usepackage{newtxtext,newtxmath}
\usepackage[T1]{fontenc}
\usepackage{ae,aecompl}

\usepackage{graphicx}	% Including figure files
\usepackage{amsmath}	% Advanced maths commands
\title[Giant HH flow in B35]{HH 175: A Giant HH Flow Emanating From A Multiple Protostar }

\author[Bo Reipurth \& Per Friberg]{
Bo Reipurth$^{1}$\thanks{E-mail: reipurth@hawaii.edu}
and Per Friberg$^{2}$
\\
$^{1}$ Institute for Astronomy, University of Hawaii at Manoa, 
640 N. Aohoku Place, Hilo, HI 96720, USA \\
$^{2}$ James Clark Maxwell Telescope, East-Asian Observatory, 660 North Aohoku Place, Hilo, HI 96720, USA}

\date{Accepted XXX. Received YYY; in original form ZZZ}

\pubyear{2021}

\begin{document}
\label{firstpage}
\pagerange{\pageref{firstpage}--\pageref{lastpage}}
\maketitle

\begin{abstract}
HH~175 is an isolated Herbig-Haro object seen towards the B35 cloud in
the $\lambda$~Ori region. We use deep Subaru 8m interference filter
images and Spitzer images to show that HH~175 is a terminal shock in a
large collimated outflow from the nearby embedded source
IRAS~05417+0907. The body of the eastern outflow lobe is hidden by a
dense ridge of gas. The western outflow breaks out of the front of the
cometary-shaped B35 cloud, carrying cloud fragments along, which are
optically visible due to photoionization by the massive $\lambda$~Ori
stars. The total extent of the bipolar outflow is 13.7~arcmin, which
at the adopted distance of 415~pc corresponds to a projected dimension
of 1.65~pc. The embedded source IRAS 05417+0907 is located on the flow
axis approximately midway between the two lobes, and near-infrared
images show it to be a multiple system of 6 sources, with a total
luminosity of 31 L$_\odot$. Millimeter maps in CO, $^{13}$CO, and
C$^{18}$O show that the B35 cloud is highly structured with multiple cores,
of which the one that spawned IRAS 05417+0907 is located at the apex
of B35. It is likely that the embedded source is the result of
compression by an ionization-shock front driven by the $\lambda$~Ori
OB stars. 
\end{abstract}


%%%%%%%%%%%%%%%%%%%%%%%%%%%%%%%%%%%%%%%%%%%%%%%%%%

% by \citet{Others2013} -  authors aim to solve \citep[e.g.][]{Author2012}.

%%%%%%%%%%%%%%%%% BODY OF PAPER %%%%%%%%%%%%%%%%%%

\section{Introduction}  \label{sec:intro}

In the mid- to late 1990's, with the advent of large-format CCDs, it
was recognized that Herbig-Haro (HH) objects can be part of giant
shocked outflow structures extending on a scale of one to several
parsec (e.g., Reipurth et al. 1997a).  Of the more than 1000 HH flows
known, several dozen have been identified as having parsec-scale
dimensions. For reviews on HH objects, see Reipurth \& Bally 2001,
Bally 2016). We here present the discovery and study of a new giant HH
flow located in the Barnard~35 (B35) cloud in the $\lambda$~Ori star
forming region.

$\lambda$ Orionis is an O8III star (Conti \& Leep 1974) that is part
of a small group of OB stars (Murdin \& Penston 1977) which excites
the HII region Sh2-264 (Sharpless 1959, Sahan \& Haffner 2016).
Bordering and bounding the HII region (which has an electron density
of about 2 cm$^{-3}$ and a mass of about $5 \times 10^3 M_{\odot}$) is
a large ring or shell of molecular clouds, detected in the optical and
in neutral hydrogen (Wade 1957; Heiles \& Habing 1974, Sahan \&
Haffner 2016), in molecular gas (Maddalena et al. 1986, Maddalena and
Morris 1987), and at mid- and far-infrared wavelengths (Zhang et
al. 1989).  The shell is highly structured, with the most massive part
in the north-west section.  The mass of HI in the ring has been
estimated to be about $4.5 \times 10^4 M_{\odot}$ (Wade 1957).
Maddalena \& Morris
(1987) adopt a value of about $1 \times 10^4 M_{\odot}$ for HI and
about $3 \times 10^4 M_{\odot}$ for H$_2$ associated with the
remaining molecular clouds.  Thus the total mass of gas in the ring is
about $4 \times 10^4 M_{\odot}$ (this number has at least a factor of
2 uncertainty).  The original molecular cloud that gave birth to this
small OB association must have had a mass of order $5 \times 10^4
M_{\odot}$, typical for a small GMC.  The dynamical age of the
expanding ring is about $5 \times 10^6$ years, in agreement with
estimates of the age of the $\lambda$-Ori OB sub-group.

$\lambda$~Ori is the most massive member of a loose cluster,
Collinder~69, which contains several hundred young low-mass stars,
detected by their H$\alpha$ emission (Joy 1949, Haro et al. 1953,
Manova 1959, Duerr et al. 1982, and Dolan \& Mathieu 2001), by X-ray
surveys (Barrado et al. 2011, Franciosini \& Sacco 2011), and infrared
surveys (Bouy et al. 2009, Hern\'andez et al. 2010).  An extensive and
unbiased spectroscopically confirmed census of the low-mass stars and
brown dwarfs in the region is given by Bayo et al. (2011).  Although
some of these stars may have been born at the same time as the O and B
stars, studies of other similar groups suggest that some of the stars
are second-generation and only a few million years old. The densest
subclusters of young stars are found near $\lambda$~Ori itself, and in
front of the small dense cloud Barnard~35. 

\begin{figure*}
%\vspace{2cm}
%\epsscale{0.8}
\includegraphics[angle=0,width=12cm]{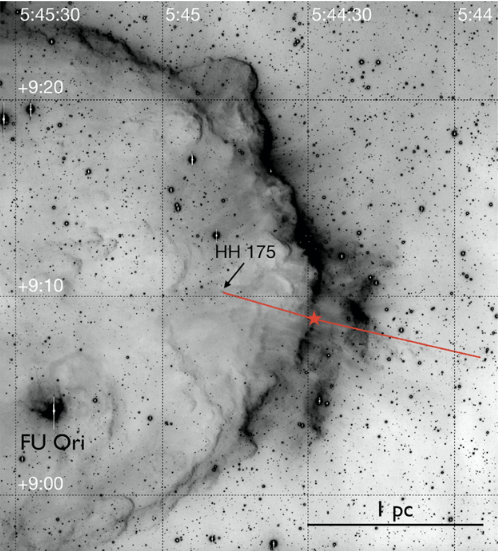}
%\vspace{0.5cm}
\caption{The cometary cloud B35 is facing the O-star $\lambda$~Ori, and consequently has a prominent bright rim. The giant HH~175 flow is indicated together with its driving source. H$\alpha$ image obtained with the Subaru telescope. 
\label{fig:subaru}}
\end{figure*}

The B35 molecular cloud (aka Lynds 1594) lies towards the projected
interior of the $\lambda$-Ori ring and shows a pronounced cometary
shape as seen in Figure~\ref{fig:subaru}. The ``tail'' of the comet
extends for over a degree to the east of the B35 cloud core (Lada \&
Black 1976), directly away from the center of the $\lambda$-Ori OB
subgroup.  This morphology provides strong evidence for direct
interaction between radiation fields and/or expansion of the HII
region excited by the massive stars in this region. The famous star FU
Orionis is located in the south-eastern wing of the B35 cloud,
surrounded by a large reflection nebula (e.g., Herbig 1966). Early
attempts to detect an embedded population at near- and far-infrared
wavelengths were not successful (Lada \& Wilking 1980, Lada et
al. 1981). An overview of the literature on the $\lambda$~Ori region is given by Mathieu (2008).

\vspace{-0.7cm}

%Figure 1

\section{Observations}  \label{sec:obs}

The B35 cloud was imaged with SuprimeCam on the Subaru 8m telescope at
Mauna Kea under program 2005b-UH-53A. We used an H$\alpha$ filter
(N-A-L659) on Jan 4, 2006 and the exposure was 5$\times$6 min in a
seeing of about 0.95 arcsec. The [SII] image was obtained with the
N-A-L671 filter on Jan 5, 2006 and the exposure was 5$\times$6 min in
a seeing of about 0.55 arcsec. The weather was clear. 

JCMT observations were obtained under the University of Hawaii program
m18bh13a and m19a16. Searches of the JCMT science archive generated
additional data from 2007, 2010 and 2017. These data were also included
in the reduction, but the bulk of the data came from 2018 and 2019.

The CO, $^{13}$CO and C$^{18}$O J=3-2 observations of B35 were
obtained using the HARP array receiver and the auto-correlator ACSIS
(Buckle et al. 2009). A 900$\times$900 arcsecond area was mapped in CO
J=3-2 using raster scanning. The line was observed with ACSIS using a
bandwith of 250 MHz and a resolution of 30.5 kHz or 0.026 km/s. This
resolution has been smoothed to improve the noise level as
required. $^{13}$CO and C$^{18}$O J=3-2 was observed at the same time
by splitting the ACSIS into two 250 MHz sub-bands with a spectral
resolution of 61 kHz or 0.052 km/s. Due to the higher opacity and
weaker emission at $^{13}$CO and C$^{18}$O an area of 300$\times$210
arcsecond was mapped covering the central source area. Some of the
archived data covered the full 900$\times$900 arcsecond area. The data
reduction was performed with the starlink {\em smurf} software
(Jenness et al. 2008).

The SCUBA2 observations were 900 arcsec {\em pong} observations at 850
and 450~$\mu$m (Holland et al. 2013). In addition a small amount of
SCUBA2 data covering the central source was obtained from the
archive. The data were reduced using the starlink {\em smurf}
reduction package (Chapin et al. 2013). In addition to IRAS
05417+0907, only two other sources were detected at 850~$\mu$m, namely the young stars V629~Ori and QR~Ori.

\section{RESULTS} \label{sec:results}

\subsection{Distance} \label{subsec:dist}

In order to properly estimate physical properties of the HH~175 flow,
the distance to B35 is important. Murdin \& Penston (1977) derived a
distance of 440 $\pm$ 40~pc based on main-sequence fitting of 11
early-type stars in the $\lambda$~Ori region. Subsequently, Dolan \&
Mathieu (2001) suggested a distance of 450 $\pm$ 50~pc based on
Str\"omgren photometry of OB stars in the region. 

More recently, Zucker et al. (2019,2020) used Gaia distances of stars
in front of and behind the $\lambda$~Ori cloud, and determined a mean
distance for seven sightlines of 410~$\pm$20~pc.

We have obtained Gaia DR2 parallaxes for the 20 stars around
$\lambda$~Ori listed in Table~5 of Dolan \& Mathieu (2001). One of
these stars, HD 36267, is a binary with distances of 108 and 110~pc,
and another, HD~35729, has a distance of 187~pc. We assume these 3
stars are foreground objects and hence reject them. By inverting the
parallaxes of the remaining stars we derive a mean distance of 420~pc.
From the cometary morphology of B35, it appears that it is seen from
the side facing $\lambda$~Ori, and we assume it is at the same
distance as the central OB stars.

We have also obtained the Gaia distance to the nearby FU~Ori,
the brightest young star in B35, and find a distance of
416$^{+9}_{-8}$~pc, in excellent agreement with the above
estimates. In the following we adopt a distance to B35 of 415~pc.

\begin{figure}
\begin{center}
%\includegraphics[width=\columnwidth]{fig-hh175-identifications.png}
% USE fig-hh175-identifications.png FOR FINAL SUBMISSION
\includegraphics[width=\columnwidth]{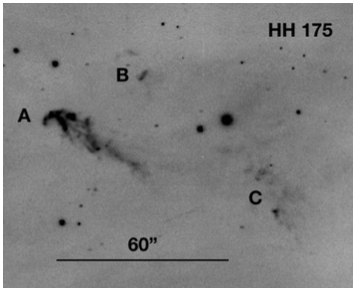}
\vspace{-0.2cm}
\caption{The HH~175 complex at the head of the eastern lobe of the outflow, as seen in a [SII] image. The main structures are labeled. The apex of the bright shock A is at 5:44:48.3 +9:10:15 (2000). \label{fig:id}}
\end{center}
\end{figure}
% Figure 2

\subsection{HH 175: Optical Imaging and Spitzer Data} \label{subsec:imaging}

HH 175 was discovered by Reipurth (1999). It has an elongated
structure with a bright head, and several more knots adjacent to it,
which are labeled in Figure~\ref{fig:id}. Interference filter
images obtained in H$\alpha$ and [SII] filters show that the object is
weak in H$\alpha$ but strong in [SII], indicating that the shock is
weak. The morphology of HH~175 suggests that its driving source may
lie to the WSW.

\begin{figure}
\vspace{0.5cm}
%\epsscale{0.9}
\includegraphics[width=\columnwidth]{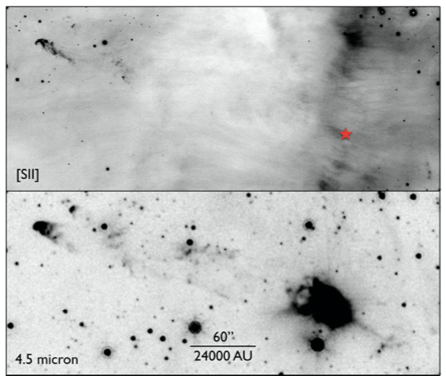}
\caption{The eastern half of the HH~175 flow as seen in the optical ([SII]) and infrared (4.5~$\mu$m). The red asterisk indicates the source location.
\label{fig:optical-ir}}
%\end{figure}
%Figure 3

%\begin{figure}
%\epsscale{0.9}
% REPLACE WITH ORIGINAL-SIZED FIGURE
\includegraphics[width=\columnwidth]{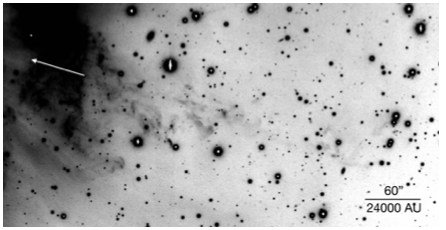}
\caption{The western part of the HH~175 flow 
where it has burst out through the bright rim of the B35 cloud, as seen in the optical (H$\alpha$). The multitude of scattered knots are designated as HH175W.
The white arrow points to the source.  
\label{fig:hh175west}}
\end{figure}

Figure~\ref{fig:optical-ir} shows the region from HH~175 to the WSW in
the optical and at 4.5~$\mu$m as observed by Spitzer.  The emission
seen in the IRAC bands is mainly from thermal molecular hydrogen at
non-LTE, although in cases where the 4.5~$\mu$m emission is stronger
(as appears to be the case for HH~175), additional emission is
required, most probably from CO vibrational emission (e.g., Takami et
al. 2010). It is immediately evident that the 4.5~$\mu$m emission can
be seen essentially all the way from the apex of HH~175 to a bright
infrared source identified as IRAS~05417+0907, which is deeply
embedded and not detected at optical wavelengths. The flow appears as
a long tube that at its widest is about 40~arcsec ($\sim$16,000~AU)
wide. Towards the apex, the flow breaks up and two small separate
fingers are seen, reminiscent of the (much more) broken-up HH~2
working surface (e.g., Herbig \& Jones 1981). The apex of each finger
coincides with HH~175 A and B, respectively. It follows that HH~175 is
just the front of a very long outflow, about 5~arcmin long, most of
which is heavily obscured. At the distance of B35, this implies an
extent of the eastern lobe of 0.6~pc. Assuming that the HH flow moves
with a tangential velocity of $\sim$100~km/s, then the dynamical age
of HH~175 is about 6,000~yr.

%Figure 4

While the eastern side of the HH~175 flow is clearly seen in the
optical and infrared images, the western side is very different. Only
a little HH emission is visible immediately west of the source, one
small group of knots can be distinguished in the lower right corner of
Figure~\ref{fig:optical-ir}-top, and we denote it HH~175W. The difficulty
in identifying HH emission here is partially because the flow is
passing through the luminous bright rim of the B35 cloud with its
disturbed emission structure. But when continuing beyond the edge of
the molecular cloud, a complex of numerous knots and filaments appear,
with the whole structure pointing back towards the IRAS source
(Figure~\ref{fig:hh175west}). We interpret this as a result of the flow
bursting out of the molecular cloud and entraining and dragging gas
and dust into the ionized zone. This is further discussed in
Section~\ref{subsec:breakout}.

The overall length of the HH 175 flow from the apex of HH~175 in the
east to the most distant H$\alpha$ fragment in the western lobe is
13.7 arcmin, corresponding to a projected extent of 1.65~pc at the
distance of 415~pc.

\begin{figure}
%\epsscale{0.7}
\includegraphics[width=\columnwidth]{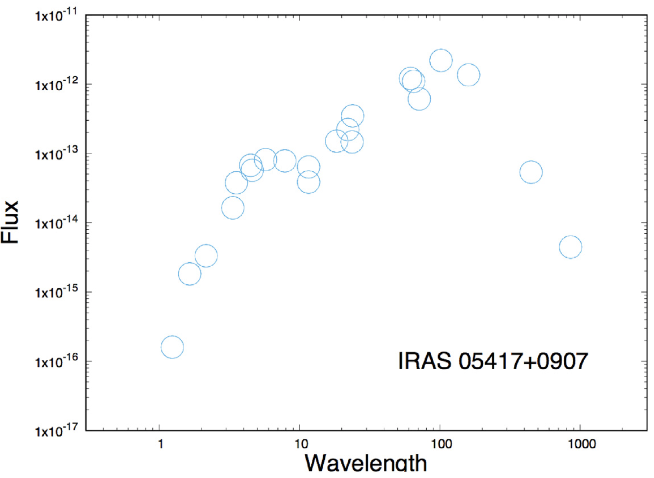}
%\vspace{-2.5cm}
\caption{The energy distribution of IRAS 05417+0907  based on data from 2MASS, Spitzer, WISE, IRAS, Akari, and SCUBA-2. The flux values are listed in Table~\ref{tab:sed}.
\label{fig:sed}}
\end{figure}
% Figure 5
% THIS FIGURE WAS MADE WITH PLOT-FILE 'ED-final2' 
% UNDER SUBDIRECTORY ENERGYDISTRIBUTION

%***************************************

%\begin{table}
%	\centering
%	\caption{Flux Values for IRAS~05417+0907}
%	\label{tab:sed}
%	\begin{tabular}{lccr} % four columns, alignment for each
%		\hline
%		$\lambda$ [$\mu$m]  & Flux [???????]    & Source \\
%		\hline
%		1.24 & 1.60 10$^{-16}$ & 2MASS J \\
%                1.65 & 1.82 10$^{-15}$ & 2MASS H \\		
%                2.16 & 3.33 10$^{-15}$ & 2MASS K \\
%                3.35 & 1.64 10$^{-14}$ & WISE W1 \\                
%                3.55 & 3.77 10$^{-14}$ & IRAC1   \\
%                4.49 & 6.74 10$^{-14}$ & IRAC2   \\
%                4.60 & 5.72 10$^{-14}$ & WISE W2 \\
%                5.73 & 8.16 10$^{-14}$ & IRAC3   \\
%                7.87 & 7.81 10$^{-14}$ & IRAC4   \\
%                11.6 & 3.86 10$^{-14}$ & WISE W3 \\
%                11.6 & 6.39 10$^{-14}$ & IRAS    \\
%                18.4 & 1.51 10$^{-13}$ & AKARI   \\
%                22.1 & 2.21 10$^{-13}$ & WISE W4 \\
%                23.7 & 1.47 10$^{-13}$ & MIPS    \\
%                23.9 & 3.49 10$^{-13}$ & IRAS    \\
%                61.8 & 1.21 10$^{-12}$ & IRAS    \\
%                65.0 & 1.10 10$^{-12}$ & AKARI   \\
%                71.4 & 6.13 10$^{-13}$ & MIPS    \\
%                102  & 2.20 10$^{-12}$ & IRAS    \\
%		160  & 1.36 10$^{-12}$ & AKARI   \\
%                447  & 5.36 10$^{-14}$ & SCUBA2  \\
%                855  & 4.45 10$^{-15}$ & SCUBA2  \\
%		\hline
%	\end{tabular}
%\end{table}

%***************************************

\subsection{The HH 175 Source} \label{subsec:source}

\begin{figure*}
\begin{center}
\includegraphics[angle=0,width=16cm]{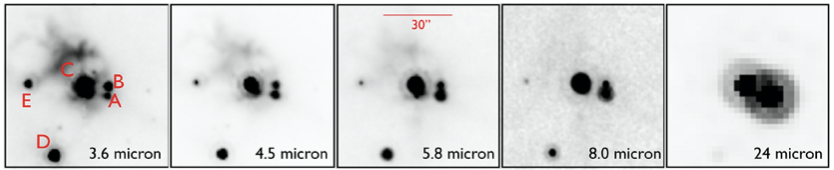}
%\vspace{-4.2cm}
\caption{Spitzer images of the source.  Each panel is slightly larger than one arcmin wide. Source~A is the dominant source at long wavelengths. 
\label{fig:spitzer}}
\end{center}
\end{figure*}

In Figure~\ref{fig:sed} we show the spectral energy distribution (SED) of
the source IRAS~05417+0907 based on observations from PanSTARRS,
2MASS, Spitzer, WISE, IRAS, Akari, and our own SCUBA2 sub-mm data. The
SED rises up to a peak around 100~$\mu$m, indicating an envelope with
a temperature around 30~K. A clear silicate dust absorption feature is
seen around 10~$\mu$m. IRAS~05417+0907 is a low-luminosity source,
Morgan et al. (2008) suggested a L$_{bol}$$\sim$6 calculated from a
blackbody fit to IRAS and SCUBA2 450 and 850~$\mu$m. The more detailed
SED presented here yields a luminosity of 15 L$_\odot$ at the adopted
distance of 415~pc.  The luminosity and SED indicate that the object
is a low-mass Class~I protostar. The object is also listed as 2MASS
J05443000+0908573 with H=15.91 and K=12.40. We have examined archival
Spitzer images and in Figure~\ref{fig:spitzer} show the source and its
environment from 3.6 to 24~$\mu$m. It is readily seen that the source
is not a single object, but forms a small multiple system. We have
labeled the various components A-E, and note that whereas the 2MASS
source (labeled C) is the dominant source at 3.6~$\mu$m, another
source to the WSW increases in brightness at longer wavelengths, until
it dominates the group at 24~$\mu$m. The above luminosity
determination is likely to include several of these sources. The
coordinates of the 5 sources are listed in Table~\ref{tab:coor}.

%====================MNRAS:

\begin{table}
   \centering
\caption{Coordinates of IRAS 05417+0907 Trapezium Sources (from Spitzer 3.5~$\mu$m image)} 
      \label{tab:coor}    
\begin{tabular}{lrr}
    \hline
     Source & $\alpha$$_{2000}$ & $\delta$$_{2000}$ \\			
    \hline      
A   &  5:44:29.27    &  +09:08:52.6     \\       
B   &  5:44:29.24    &  +09:08:56.7     \\            
C1  &  5:44:29.96    &  +09:08:57.0     \\           
C2  &  5:44:29.92    &  +09:08:55.9     \\          
D   &  5:44:30.86    &  +09:08:26.3     \\           
E   &  5:44:31.64    &  +09:08:57.9     \\
    \hline
\end{tabular}
\end{table}

%Figure 6

\begin{figure}
\begin{center}
\includegraphics[angle=0,width=\columnwidth]{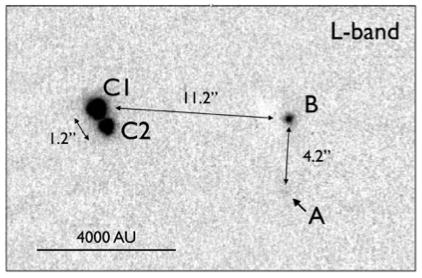}
\vspace{-0.3cm}
\caption{An L-band image of IRAS~05417+0907 showing the components
of the multiple system. The source A, which dominates at longer
wavelengths, is barely detected in this exposure which was kept short
to not saturate the bright binary. Image from Connelley et al. (2008). 
\label{fig:multiplicity}}
\end{center}
\end{figure}

%Figure 7

\begin{figure*}
\begin{center}
%\vspace{-1cm}
%\includegraphics[angle=0,width=16cm]{fig-Spitzerpanel2-reduced.png}
\includegraphics[angle=0,width=16cm]{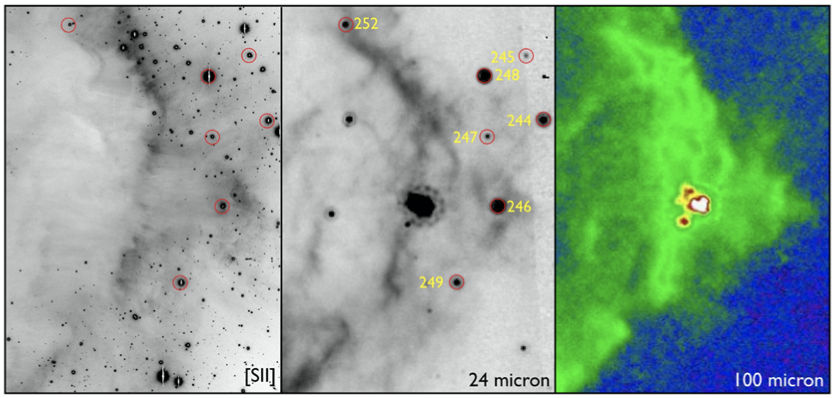}
%\vspace{-2.0cm}
\caption{A multi-wavelength panel featuring the apex of the B35 cloud in a
Subaru [SII] filter, and Spitzer MIPS 24~$\mu$m and Herschel 100
$\mu$m maps. The center image has known pre-main sequence
stars marked, labeled following the catalog of Dolan \& Mathieu
(2001). The Herschel panel shows the small IRAS 05417+0907 trapezium,
revealing that it is currently the only protostars embedded in the
cloud. Each figure is 6 arcmin wide.
\label{fig:triptych}}
\end{center}
\end{figure*}

%Figure 8

\begin{figure}
%\epsscale{0.5}
\includegraphics[width=\columnwidth]{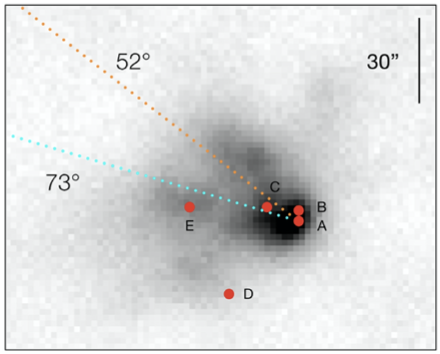}
%\vspace{-2.3cm}
\caption{The HH 175 driving source IRAS~05417+0907 observed at 450~$\mu$m with SCUBA-2. What appears to be the dusty walls of an outflow cavity is opening away from the source. The direction to the HH~175 bow shock is indicated as a turquoise dotted line, and the direction to the peak intensity of the molecular outflow is marked in orange.  
\label{fig:submm}}
\end{figure}

%Figure 9

\begin{figure*}
%\vspace{2cm} 
\begin{center}
%\includegraphics[angle=270,width=18cm]{fig-mosaic.pdf}
% USE fig-mosaic.pdf IN THE FINAL SUBMISSION
\includegraphics[angle=0,width=18cm]{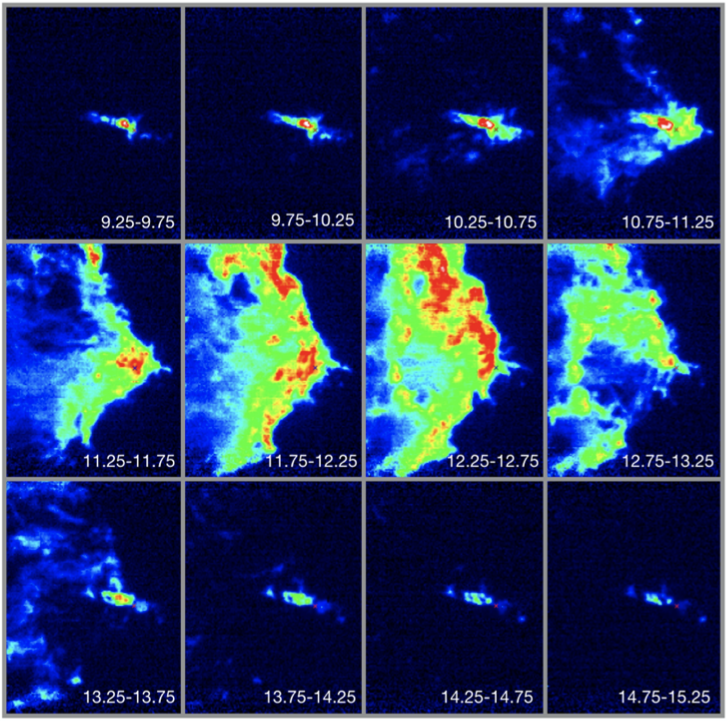}
\vspace{0.3cm} 
\caption{A mosaic of 12 panels showing B35 and the HH~175 molecular outflow in $^{12}$CO from blueshifted (top left) to redshifted (bottom right). The v$_{lsr}$ velocities spanned by each panel is indicated. The panels are 18.4~arcmin high, corresponding to 2.25~pc. 
\label{fig:mosaic}} 
\end{center}
\end{figure*}

%Figure 10

\begin{figure}
%\epsscale{0.7}
%\includegraphics[width=\columnwidth]{fig-contours-CO-annot2.png}
% USE fig-contours-CO-annot2.png FOR FINAL SUBMISSION!
\includegraphics[width=\columnwidth]{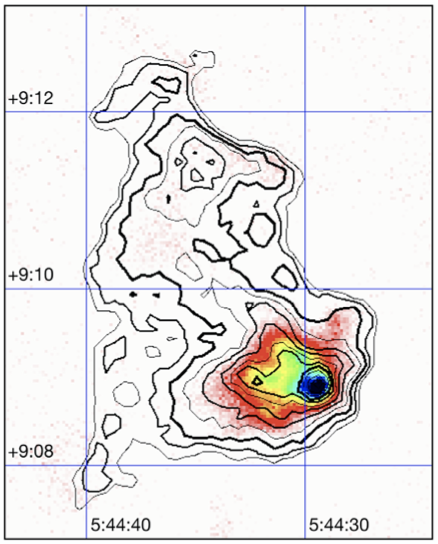}
%\vspace{-2cm}
\caption{The structure of the B35 cloud observed in C$^{18}$O
overlaid on a color-map of the same region observed at 450$\mu$m with
SCUBA2 on JCMT. It is evident that the little cluster around IRAS 05417+0907 is the only actively star forming region in B35.
\label{fig:contour}}
\end{figure}

%Figure 11

\begin{figure*} 
%\vspace{1cm} 
\begin{center}
\includegraphics[angle=0,width=14cm]{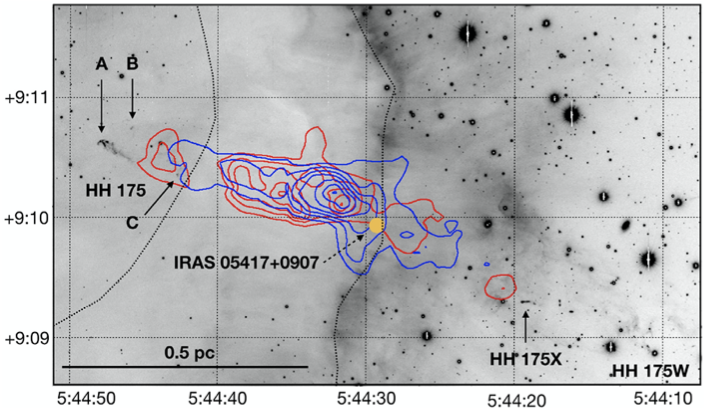}
%\vspace{-0.8cm} 
\caption{The molecular outflow driven by IRAS
05417+0907 plotted on a [SII] optical image. The eastern lobe is
prominent where the flow burrows through the cloud, whereas the
western lobe is much less prominent as it breaks out of the
cloud. The dense ridge observed in CO is indicated by dotted lines. 
\label{fig:flow}} 
\end{center}
\end{figure*}

%Figure 12

\begin{figure}
%\epsscale{0.6}
\includegraphics[angle=0,width=\columnwidth]{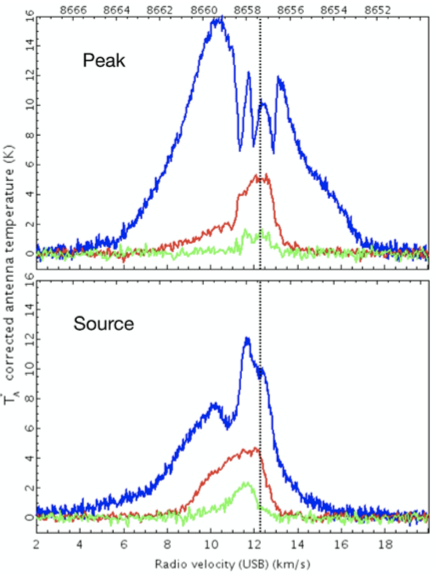}
\vspace{-0.5cm}
\caption{Line profiles observed with JCMT towards two regions in
the HH~175 molecular outflow. On top is the pointing towards the region
with strongest emission   
 (5:44:31.8 +9:09:19 - 2000)
just ENE of the source (see
Figure~\ref{fig:flow}), and below is the emission towards the source 
(5:44:29.8 +9:08:54 - 2000). Blue
is $^{12}$CO J=3-2, red is $^{13}$CO J=3-2, and C$^{18}$O J=3-2 is green. 
The dotted
line indicates the mean rest-velocity 12.25~km/sec of the B35 cloud.
\label{fig:combi}}
\end{figure}

%Figure 13

\begin{figure}
 \begin{center}
\includegraphics[angle=0,width=\columnwidth]{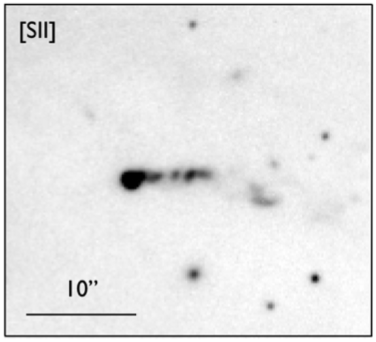}
%\vspace{0.5cm} 
\caption{HH~175X as seen in a [SII] image from the Subaru
telescope. HH~175X is located within the western lobe
of the giant HH~175 flow, and may be a shocked fragment of the outflow, but because of its collimated morphology and unusual brightneww it cannot be excluded that it is an independent jet. The star at the eastern end of HH~175X is at 
5:44:19.5 +09:07:35 (2000).    
\label{fig:hh175X}} 
\end{center}
\end{figure}

%Figure 14

Connelley et al. (2008) obtained a L-band image of IRAS 05417+0907 in
a survey for protostellar binaries, and they found source C to be a
close binary with a separation of 1.2~arcsec.
Figure~\ref{fig:multiplicity} shows an L-band image of the source region,
based on data from Connelley et al. (2008). The exposure is brief in
order to not saturate the brightest source. It is seen that IRAS
05417+0907 is a quadruple system, arranged in a non-hierarchical
configuration. As mentioned above, the source detected by IRAS is
source A, which is barely detected in this short exposure.

Herschel observed the region, and Figure~\ref{fig:triptych} shows a
comparison of the apex of the B35 cloud as imaged with Subaru in a
[SII] filter, with Spitzer at 24~$\mu$m and with Herschel at
100~$\mu$m. Known young stars from the Dolan \& Mathieu (2001) catalog
are marked with red circles.  Comparing the images it is evident that
IRAS 05417+0907 is currently the only protostar at the apex of the
cloud. The B35 cloud is located at the eastern edge of the HII region
around $\lambda$~Ori, and its cometary shape and bright rims testifies
to the effect of the central OB stars. It appears that IRAS~05417+0907
is another case of triggered star formation (e.g., Sugitani et
al. 1991).

IRAS~05417+0907 source A was detected at 6 cm with the VLA by Terebey
et al. (1992), who also detected two highly variable H$_2$O masers
located within a fraction of an arcsecond from the position of
source~A given in Table~1. Additional detections were made by Claussen
et al. (1996). We have observed the source region with SCUBA2 on the
JCMT at 450 and 850~$\mu$m, and found no other submm sources in the
field (Figure~\ref{fig:submm}). IRAS~05417+0907 appears as a bright
unresolved source with a conical nebula to the north-east along the
flow axis, suggesting that this is the outflow cavity illuminated and
heated by the jet.  Perotti et al. (2021) have carried out a major
near-infrared and mm-study of the environment of IRAS~05417+0907 in
order to calculate the gas-to-ice ratio for CO and CH$_3$OH, thus
casting light on how these solid-state molecules on dust grains are
converted into the gas phase.

Finally, we examine the evidence for youth of the components of the
IRAS 05417+0907 components. The detection of A, C, and D at 100~$\mu$m
combined with the IRAS colors show that they are young embedded
protostars. Additionally, Connelley et al. (2007) detected reflection
nebulae around C and D. The detection of B out to 24~$\mu$m and E out
to 8~$\mu$m are indicative but not proof that they are young.

The dynamical evolution of the IRAS 05417+0907 trapezium and its relation to
HH~175 is discussed in Section~\ref{subsec:multi}.

% (2000): 5 44 29.26, +09 08 52.4 
% (1950): 5 41 44.80, +09 07 38.7

\subsection{Cloud Structure and Molecular Outflow} \label{subsec:mm}

A low-resolution mm-study of B35 was carried out by Lada \& Black
(1976), who found that the $^{12}$CO column density peaks behind the
western ionization front. Myers et al. (1988) and Benson \& Myers
(1989) made a small low-resolution map in $^{12}$CO and NH$_3$ around
the IRAS 05417+0907 source, and discovered an energetic bipolar
molecular outflow emanating from a dense core. Additional
low-resolution $^{12}$CO observations were made by Qin \& Wu
(2003). Craigon (2015) performed a more detailed mm study of B35 and
the molecular outflow and suggested a correlation between the gas
temperature and PAH emission consistent with photoelectric heating.

Our JCMT/HARP observations form the deepest and most detailed mm study
of B35 to date, for technical details see Section~\ref{sec:obs}. A
mosaic of 12 channel maps with 8'' resolution (pixel size 6'') in
$^{12}$CO J=3-2, covering almost all of B35, is presented in
Figure~\ref{fig:mosaic}. The velocity range shown covers
9.25 $<$ v$_{lsr}$ $<$ 15.25 km/sec. The bulk of the molecular cloud
is seen between 11.75 and 12.75 km/sec, and we adopt a mean velocity
of 12.25 km/sec as the average cloud velocity. The enhancement of
$^{12}$CO emission behind the ionization-shock front is most clearly
seen in the 11.75-12.25 km/s panel.  It also appears that there is a
slight asymmetry in the velocity distribution, such that the southern
part of B35 moves towards the observer while the northern part tends
to move away. The velocity gradient is about 1~km/sec from North to
South. The C$^{18}$O J=3-2 contour map in Figure~\ref{fig:contour}
shows that the B35 cloud is highly fragmented, and that the HH~175 source is
located in the very densest part of the B35 cloud (the colored part of
the figure is from the 450~$\mu$m SCUBA2 map).

To derive a cloud mass, we assume a distance of 415~pc and a
N(H$_2$)/N(13CO) ratio of 7.1$\times$10$^5$ (Frerking et al. 1982). We
can derive a mass assuming we know the excitation temperature T$_{ex}$
and that $^{13}$CO is optically thin (Mangum \& Shirley 2015). The
excitation temperature will vary across the map and along the
line-of-sight which passes through both the warm skin of the cloud and
the colder interior regions and, additionally, $^{13}$CO is likely
partially optically thick. Without detailed knowledge of these issues,
we have chosen to adopt an excitation of 25~K along the northwestern
edge and of 16~K in the eastern interior of the cloud based on the
$^{13}$CO excitation temperature map derived by Craigon (2015). Using
these two uniform temperatures we find a mass of 36 M$_\odot$ for B35.
The uncertainty in this determination is considerable due to the above
mentioned assumptions. As a comparison we get 39 and 29 M$_\odot$ for
excitation temperatures of 16 and 25 K, respectively, assuming they
are uniform across the cloud.

Figure~\ref{fig:mosaic} very clearly shows the molecular outflow emanating
from IRAS 05417+0907 at a position angle of $\sim$72$^\circ$. As already
noted by Craigon (2015) the outflow is significantly
collimated. Figure~\ref{fig:flow} outlines the blue and red lobes overlaid
on the deep SuprimeCam image. The figure shows that the eastern lobe
is very extended, reaching almost out to HH~175. In contrast, the
western lobe is short and stubby. This is well understood when
compared with the extinction seen in the optical image: the driving
source is located at the western edge of a dense rim, through which the flow
burrows until emerging on the eastern side of the rim. As it escapes at
the cloud rim it no longer passes through dense gas to entrain, and
hence does not reach all the way to the HH object. Similarly for the
western lobe, which escapes into the tenuous ambient medium in front
of B35. The eastern lobe shows both strong blue and red emission,
indicating that the flow axis is very close to the plane of the sky,
in agreement with the results of Craigon (2015). Figure~\ref{fig:combi}
shows the line profiles of $^{12}$CO, $^{13}$CO, and C$^{18}$O towards
the peak emission and towards the source.

Due to the flow having a very low angle of inclination to the plane of
the sky, there is substantial overlap in velocity between the outflow
and the ambient cloud.  This is an obstacle to an estimate of the
outflow mass. Craigon (2015) subtracts a scaled quiescent cloud line
profile centered at 12.42~km/s to remove the ambient cloud
contribution from the flow estimates. However, looking at the spectral
lines in Figure~\ref{fig:combi}, their profiles are complex and
towards the source C$^{18}$O covers a wider interval than the ambient
cloud and is centered below 12.42 km/s.  We have instead integrated
the flow over selected velocity intervals.  The red and blue line
wings were integrated over a sequence of velocity intervals ending
closer and closer to the line center. The resulting maps were then
spatially integrated over the flow region. This was done both for the
$^{12}$CO and $^{13}$CO J=3-2 lines. The ratio between integrated
$^{12}$CO and $^{13}$CO values is about 70 in the red wing until the
lower edge in the velocity range reaches 13.5~km/s. Below 13.5~km/s
the ratio begins to drop fast. We assume this is due to contribution
from the ambient cloud and cloud core. The blue wing is partially
optically thick, the ratio between integrated $^{12}$CO and $^{13}$CO
values are about 12 until the upper edge in the velocity range reaches
11.0~km/s at which point it starts to drop. Hence we have adapted
11.0~km/s to 13.5~km/s as the region mostly affected by the ambient
cloud, and therefore exclude this when integrating the outflow over
velocity.

Outflow masses for the two lobes have been calculated for the
following regions: a rectangular area with width of 250~$\arcsec$ and
tilted 28.5 degrees following the flow area for 310~$\arcsec$ towards
east-northeast from Source~A and 140~$\arcsec$ towards west-southwest from
Source~A.  

% For an adopted excitation temperature of 25~K for $^{13}$CO
% and a (very poor) assumption that it is optically thin, the large

The flow mass has been calculated using the $^{13}$CO transition,
since $^{12}$CO is likely optically thick and the line profiles in
Figure~\ref{fig:combi} show signs of self absorption. Assuming that
the $^{13}$CO line is optically thin and adopting an excitation
temperature of 25~K, the large eastern lobe has a total mass of
0.60~M$_\odot$ (blue 0.54~M$_\odot$, red 0.06~M$_\odot$) and the
smaller western lobe has a total mass of 0.37~M$_\odot$ (blue
0.37~M$_\odot$, red 0.002~M$_\odot$). In total, we find an outflow
mass (lower limit) of $\sim$0.97~M$_\odot$, which compares well with
the mass of 0.86~M$_\odot$ derived by Myers et al. (1988). For
excitation temperatures of 35~K and 16~K, lower limits for the total
flow masses are 0.93~M$_\odot$ and 1.30~M$_\odot$, respectively. Given
the many physical and geometric uncertainties involved, it is
difficult to estimate the uncertainty of these numbers, but it appears
that the mass of the HH~175 molecular outflow is in the same range as
for many other molecular outflows (e.g., Bally et al. 1999, Lee et
al. 2002).

%+======================================
%Hi Bo,

%I have recalculated the flow masses to ensure they cover the same
%region. The calculation is for a rectangular area tilted 28.5 degrees
%to follow the flow area. The rectangle stretches 310'' towards
%east-northeast and 140'' towards west-southwest from the location of
%source A and has a width of 250''.

%                east         west       total
%red          0.06          0.002      0.06
%blue         0.54          0.37       0.91

%Note that the numbers for total (red \& blue) has changed some. If you
%change the excitation temperature to 35 (16) K the numbers are

%                east             west                   total
%red          0.05 (0.08)    0.002 (0.003)          0.052 (0.083)
%blue        0.52 (0.72)     0.36  (0.50)           0.88 (1.22)

%Per

%=====================================

% I used 35 and 16 K - the corresponding masses are 0.05 (35 K) and 0.07
% (16 K) for the red and 0.75 (35 K) and 1.04 (16 K) for the blue.

% The mass estimate is starting to go up exponetially if you go lower in
% temperature.

%===========================================

%=======================

\subsection{On the Nature of HH 175X} \label{subsec:newjet}

While examining the deep SuprimeCam images, we noted a small jet-like
feature, here named HH~175X and marked in Figure~\ref{fig:flow}. 
As can be seen in Figure~\ref{fig:hh175X}, HH~175X has the
morphology of a well collimated jet, bright in [SII] emission, with
4-5 well-defined knots stretching over 7~arcsec, corresponding to
2800~AU, and an off-axis knot further away.

Under normal circumstances we would not hesitate to identify this as
another small jet in a star forming cloud. But in this case the object
is located within the lobe of a giant HH flow bursting out through the
torn fabric of a cloud, Thus it could well be a fragment like the many
other small shocks and photoionized clumps in the outflow
lobe. Speaking against this, however are the following facts: {\em
(1)} HH~175X stands out as by far the brightest of these small scale
shocks; {\em (2)} The structure of well aligned knots would be unusal
for a random filament; {\em (3)} The jet knots appear to emerge from a
faint star; {\em (4)} The object is located within the bright rim of a
dense cloud that is known to have recently spawned numerous low-mass
stars (Dolan \& Mathieu 1999, 2001, 2002).

Each of these alone would not be sufficient to warrant much
attention, but taken together it seems at least possible that this
could be a new small HH jet.

HH~175X is not precisely aligned with the star at its eastern
end. Closer examination shows that the pointspread function of the
star differs from those of the surrounding stars, and is elongated at
a PA of about 45$^\circ$ towards the end of the jet. This suggests a
companion at $\sim$0.33~arcsec, corresponding to about 130~AU. The
coordinates of this star at the end of HH~175X are (2000): 5 44 19.56
+09 07 35.6.  The faint star is not detected by 2MASS, but is (very)
weakly seen in all four Spitzer IRAC channels. It is thus not showing
signs of significant circumstellar material.

For the time being, we think it most likely that HH~175X is another
shocked fragment in this outflow, but further observations, in
particular of proper motions, are needed to fully settle this issue.

\section{DISCUSSION} \label{sec:discussion}

The recognition and study of giant HH flows is important, because they
provide insight into a number of physical processes in star
formation. Because their dynamical time scales can be several times
10$^4$~yr, they provide a fossil record of the most recent evolution
and accretion history of their driving sources. Multi-epoch imaging
and spectroscopic analyses indicate variability in their ejection
directions and velocities. Detailed studies of HH flows indicate a
systematic decrease in their space velocities with time (e.g., Devine et
al. 1997) which is interpreted as a deceleration of the ejecta as they
penetrate the ambient medium (e.g., Cabrit \& Raga 2000). Because HH
flows transfer energy and momentum into their ambient medium, they may
be an important contribution to the maintenance of turbulence in
molecular clouds. Finally, because fast shocks will dissociate
molecules, giant HH flows can result in a chemical rejuvenation of
clouds in star forming regions.

\subsection{Observations and Models of Breakout} \label{subsec:breakout}

Parsec-scale outflows have dimensions much larger than the cloud cores
in which they originate. As a consequence, giant outflows usually
punch out of their birth clouds and inject energy, momentum, and mass
into the intercloud medium. In regions of massive star formation, this
implies that the flows and the entrained gas they drag along are
bathed in UV radiation. 

HH jets that are formed from stars located outside their nascent
clouds may, in addition to the emission from shocks, show emission
from photoionization (Cernicharo et al. 1998, Reipurth et
al. 1998). Similarly, a molecular outflow that entrains gas while
embedded and which breaks out from the neutral cloud into an HII
region will become photoionized. Bally et al. (2002) found two such
cases while studying the S140 region. S140 has a well-defined and
rather sharp interface between ionized and neutral gas and Bally et
al. found two flows, HH~616 and HH~617, bursting out of the dense
molecular gas. Subsequently, Reipurth et al. (2003) identified another
case, where an outflow, HH~777, breaks out of the IC~1396N cometary
globule. Since then, a few more cases have been found, in particular
the HST images of the HH~902 flow in Carina show the phenomenon
well (e.g. Smith et al. 2010). 
The nature of such blow-outs from cloud cores were investigated
through numerical simulations by Raga \& Reipurth (2004). By varying
physical parameters of flow and cloud core an overview of the
properties of such blowout was established. The working surfaces of
the flow essentially act as neutral clumps which are irradiated by the
impinging ionizing photons. The emission of the flows is therefore
dominated by the impinging photon field rather than by dynamical
properties of the working surfaces. Evidently, once the jet emerges
from its cloud core into the tenuous intercloud medium, it runs out of
material to entrain. It follows that the scattered fragments found in
the western lobe of HH~175 have been dragged out from the cloud
interior.  The western lobe is significantly longer than its eastern
counterpart which has penetrated the cloud, as one would expect from
momentum conservation.

For reasons that are not clear, most outflows that break out of their
birth clouds do not show evidence of such fragments dragged out of the
cloud. An example is the HH~111 jet, whose eastern lobe plunges
through the cloud, while the western lobe flows directly from the
cloud core into a void, without any evidence of swept-up cloud
material.

\subsection{Giant HH Flows and Multiple Energy Sources} \label{subsec:multi}

A very high fraction of multiplicity in the sources of giant HH flows
was noted by Reipurth (2000), and many more giant flows with multiple
driving sources have been noted by now. It is possible that there is a
causal link between multiplicity and the giant HH flows. It is well
known that non-hierarchical triple systems are unstable, and following
a close triple encounter during which energy and momentum can be
exchanged, this leads either to the formation of a hierarchical triple
system with a close binary and a distant third component, or the third
body is escaping, leaving behind a binary (Valtonen \& Mikkola 1991).

Most stars are born in small multiple systems, which decay (e.g.,
Sterzik \& Durisen 1998). In case the body that escapes has not
had time to gain sufficient mass to achieve hydrogen burning, then it
remains a brown dwarf (Reipurth \& Clarke 2001).

The binary that remains after a triple decay becomes highly eccentric
and is tightened in the process. If it furthermore moves in a
dissipative gas environment, it may spiral in, leading to a
spectroscopic binary or even a merger (Bate 2002, Reipurth et
al. 2014).

These dynamical processes lead to significant disk-disk interactions
at periastron, and result in accretion events with a consequent burst
of outflow activity (e.g., Tofflemire et al. 2017). We can thus
understand the formation of giant HH flows as a consequence of the
evolution of a newly formed binary: when a newly born triple system
undergoes a close triple encounter during which their disks violently
collide, then a major outflow event is initiated. The resulting binary
will additionally produce accretion/outflow episodes during subsequent
periastron passages, which will become increasingly frequent as the
binary spirals in. Eventually the stars are so close, of the order of
$\sim$10~AU or so, that a jet is formed with multiple closely spaced
knots. 

In the case of HH~175, we clearly see that the source forms a small
multiple system. In accordance with the scenario outlined above, we
thus postulate that source~A is now a close binary in the process of
spiraling in, following a dynamical interaction with one or more of
the other stars in the system about 6,000~yr ago, the estimated
dynamical age of HH~175.  The mean terminal velocity of ejectees from a
low-mass triple system is 1.1~km/sec (Reipurth et al. 2010),
indicating that an ejected companion would be of the order of a few
arcseconds away from source~A. The separation of source~A and B is
4~arcsec, suggesting a possible connection to the event that formed
the HH~175 flow.

\section{CONCLUSIONS} \label{sec:conclusions}

We have studied the region of the newly discovered HH~175 object in
the B35 cloud associated with the $\lambda$ Ori region, and have
reached the following conclusions:

1. Wide-field optical images together with Spitzer images reveal a
   giant outflow with an extent of 1.65~pc, in which HH~175 is the
   terminal shock of the eastern lobe, whereas the western lobe breaks
   out of the face of the B35 cloud.  For an assumed tangential
   velocity of 100~km/sec, the dynamical age of the outflow is
   $\sim$6,000~yr.

2. The driving source of the HH~175 flow is IRAS 05417+0907, an
   embedded Class~I source which in Spitzer images resolve into a
   multiple system with at least 6 components. 

3. Our $^{12}$CO J=3-2, $^{13}$CO J=3-2, and C$^{18}$O J=3-2 maps of
   the entire B35 cloud show that the IRAS source is embedded in a
   large dense cloud core and that the B35 cloud is highly fragmented.

4. The $^{12}$CO map reveals a major molecular outflow coinciding with
   the optical/IR flow. Both the eastern and western lobes show blue
   and red high-velocity wings, indicating that the outflow lies
   almost in the plane of the sky. 

5. The HH~175 flow adds to the increasing number of multiple systems
   found to drive a giant HH flow. We argue that the HH~175 giant flow
   is the result of chaotic motions in an unstable non-hierarchical
   newborn stellar system, during which close periastron passages lead
   to major disk disturbances that lead to accretion events which
   again drive strong outbursts of outflow activity.

\section*{Acknowledgements}

We thank Giulia Perotti and Helen Fraser for drawing our attention to
the thesis by Alison Craigon, and Michael Connelley for providing the
image in Figure~\ref{fig:multiplicity}. 
 We also thank an anonymous referee for helpful comments. 
 Based in part on data collected
at the Subaru Telescope, which is operated by the National
Astronomical Observatory of Japan (NAOJ).  Thanks are due to the
Subaru staff, in particular Miki Ishii and Hisanori Furusawa
(SuprimeCam) for excellent and dedicated support during the
observations.  We are grateful to Nobunari Kashikawa for permission to
use his [SII] filter.
The James Clerk Maxwell Telescope is operated by the East Asian
Observatory on behalf of The National Astronomical Observatory of
Japan; Academia Sinica Institute of Astronomy and Astrophysics; the
Korea Astronomy and Space Science Institute; the Operation,
Maintenance and Upgrading Fund for Astronomical Telescopes and
Facility Instruments, budgeted from the Ministry of Finance (MOF) of
China and administrated by the Chinese Academy of Sciences (CAS), as
well as the National Key R\&D Program of China
(No. 2017YFA0402700). Additional funding support is provided by the
Science and Technology Facilities Council of the United Kingdom and
participating universities in the United Kingdom and Canada. 
Program ID M18BH13A.
The JCMT archive is hosted by the Canadian Astronomy Data Center. 
This work is based in part on observations made with the Spitzer Space
Telescope, which is operated by the Jet Propulsion Laboratory,
California Institute of Technology under a contract with NASA
This paper has used archival data from the Herschel mission. Herschel
is an ESA space observatory with science instruments provided by
European-led Principal Investigator consortia and with important
participation from NASA.
This research has made use of the SIMBAD database, operated at CDS,
Strasbourg, France, and of NASA's Astrophysics Data System
Bibliographic Services.

\section*{Data Availability Statement}

The data underlying this article will be shared on reasonable request 
to the corresponding author.

\clearpage

\appendix

\section{Energy Distribution}

The following table lists the fluxes for
the photometric measurements shown in Figure~\ref{fig:sed}.

\begin{table}
	\centering
	\caption{Flux Values for IRAS~05417+0907}
	\label{tab:sed}
	\begin{tabular}{lcl}
		\hline
		$\lambda$ [$\mu$m]  & $\nu$F$_\nu$ [Wm$^{-2}$]    & Source \\
		\hline
		1.24 & 1.60 10$^{-16}$ & 2MASS J \\
                1.65 & 1.82 10$^{-15}$ & 2MASS H \\		
                2.16 & 3.33 10$^{-15}$ & 2MASS K \\
                3.35 & 1.64 10$^{-14}$ & WISE W1 \\                
                3.55 & 3.77 10$^{-14}$ & IRAC1   \\
                4.49 & 6.74 10$^{-14}$ & IRAC2   \\
                4.60 & 5.72 10$^{-14}$ & WISE W2 \\
                5.73 & 8.16 10$^{-14}$ & IRAC3   \\
                7.87 & 7.81 10$^{-14}$ & IRAC4   \\
                11.6 & 3.86 10$^{-14}$ & WISE W3 \\
                11.6 & 6.39 10$^{-14}$ & IRAS    \\
                18.4 & 1.51 10$^{-13}$ & AKARI   \\
                22.1 & 2.21 10$^{-13}$ & WISE W4 \\
                23.7 & 1.47 10$^{-13}$ & MIPS    \\
                23.9 & 3.49 10$^{-13}$ & IRAS    \\
                61.8 & 1.21 10$^{-12}$ & IRAS    \\
                65.0 & 1.10 10$^{-12}$ & AKARI   \\
                71.4 & 6.13 10$^{-13}$ & MIPS    \\
                102  & 2.20 10$^{-12}$ & IRAS    \\
		160  & 1.36 10$^{-12}$ & AKARI   \\
                447  & 5.36 10$^{-14}$ & SCUBA2  \\
                855  & 4.45 10$^{-15}$ & SCUBA2  \\
		\hline
	\end{tabular}
\end{table}

\end{document}